\input phyzzx.tex
\tolerance=1000
\voffset=-0.0cm
\hoffset=0.7cm
\sequentialequations
\def\rl{\rightline}

\def\t1{{\tilde 1}}

\def\t{\theta}

\REF{\COS}{S. Perlmutter et al, Astrophys. J. {\bf 483}, 565 (1997), astro-ph/9608192.}
\REF{\RAT}{B. Ratra and P. J. Peebles, Phys. Rev. {\bf D37}, 3406, (1988); R. R. Caldwell, R. Dave and P. J. Steinhardt, Phys. Rev. Lett. {\bf80} (1998) 1582.}
\REF{\ACC}{S. Hellerman, N. Kaloper and L. Susskind, hep-th/0104180; W. Fischler, A. Kashani-Poor, R. McNees and S. Paban, hep-th/0104181; X. He, astro-ph/0105005.}
\REF{\QUI}{E. Halyo, hep-ph/0105216; J. M. Moffat, hep-th/0105017; J. Cline, hep-ph/0105251; M. Li, W. Lin, X. Zhand and R. Brandenberger, hep-ph/0107160.}
\REF{\MAL}{J. Maldacena and A. Strominger, JHEP {\bf 9802} (1998) 014, gr-qc/9801096.}
\REF{\HAW}{S. Hawking, J. Maldacena and A. Strominger, JHEP {\bf 0105} (2001), hep-th/0002145.}
\REF{\WIT}{E. Witten, ``Quantum Gravity in De Sitter Space'', Talk at the Strings 2000 Conference, http://theory.tifr.res.in/strings/;  hep-th/0106109.}
\REF{\STR}{A. Strominger, hep-th/0106113.}
\REF{\LIN}{F. lin and Y. Wu, Phys. Lett. {\bf B453} (1999) 222, hep-th/9901147.}
\REF{\VJ}{V. Balasubramanian, P. Horava and D. Minic, hep-th/0103171.}
\REF{\HOL}{G. 't Hooft, gr-qc/9310026; L. Susskind, J. Math. Phys. {\bf 36} (1995) 6377, hep-th/9409089.}
\REF{\RAP}{R. Bousso, JHEP {\bf 9907} (1999) 004, hep-th/9905177; JHEP {\bf 9906} (1999) 028, hep-th/9906022; JHEP {\bf 0104} (2001) 035, hep-th/0012052.}
\REF{\LEN}{L. Susskind, hep-th/9309145.}
\REF{\SBH}{E. Halyo, A. Rajaraman and L. Susskind, Phys. Lett. {\bf B392} (1997) 319, hep-th/9605112.}
\REF{\HP}{G. Horowitz and J. Polchinski, Phys. rev. {\bf D55} (19997) 6189, hep-th/9612146.}
\REF{\DAM}{T. Damour and G. Veneziano, Nucl. Phys. {\bf B568} (2000) 93, hep-th/9907030.}
\REF{\JUA}{J. Maldacena, Nucl. Phys. {\bf B477} (1996) 168, hep-th/9605016.}
\REF{\EDI}{E. Halyo, Int. Journ. Mod. Phys. {\bf A14} (1999) 3831, hep-th/9610068; Mod. Phys. Lett. {\bf A13} (1998), hep-th/9611175.}
\REF{\DAN}{U. Danielsson, A. Guijosa and M. Kruczenski, hep-th/0106201.}
\REF{\HRS}{E. Halyo, B. Kol, A. Rajaraman and L. Susskind, Phys. Lett. {\bf B401} (1997) 15, hep-th/9609075.}
\REF{\NAR}{H. Nariai, Sci. Rep. Tohuku Univ. {\bf 34} (1950) 160; {\it ibid.} {\bf 35} (1951) 62.}

\singlespace
\rl{SU-ITP-01-31}
\rl{hep-ph/0107169}
\rl{\today}
\pagenumber=0
\normalspace
\medskip
\bigskip
\titlestyle{\bf{De Sitter Entropy and Strings}}
\smallskip
\author{ Edi Halyo{\footnote*{e--mail address: vhalyo@.stanford.edu}}}
\smallskip
\centerline {Department of Physics}
\centerline{Stanford University}
\centerline {Stanford, CA 94305}
\centerline{and}
 \centerline{California Institute for Physics and Astrophysics}
\centerline{366 Cambridge St.}
\centerline{Palo Alto, CA 94306}
\smallskip
\vskip 2 cm
\titlestyle{\bf ABSTRACT}

We show that the entropy of de Sitter space in any dimension can be understood as the entropy of a highly excited string located near the
horizon. The string tension is renormalized to $T \sim \Lambda$ due to the large gravitational redshift near the horizon.
The de Sitter temperature is given by the Hagedorn temperature of the string.

\singlespace
\vskip 0.5cm
\endpage
\normalspace

\centerline{\bf 1. Introduction}
\medskip

Recent observations suggest that the expansion of the universe is accelerating [\COS] which means that the energy of the universe is dominated by a
component with negative pressure such as a cosmological constant or quintessence[\RAT]. In either case, in the far future the universe will look like
de Sitter space which will expand exponentially. It has been argued that the formulation of string theory in these situations is problematic[\ACC]. (For ways
of getting around this result see [\QUI].)
It is well--known that an observer in de Sitter space sees a cosmological horizon and therefore there is a finite entropy and temperature in
de Sitter space. The origin of this entropy and temperature is not clear (For some of the attempts for an explanation see [\MAL-\VJ].). However, holography
[\HOL,\RAP] implies
that the essential degrees of freedom that result in the entropy of de Sitter space reside on or near the horizon (similar to the boundary of the
well--understood anti--de Sitter case). Supersymmetry cannot be defined on de Sitter space so the physics is perhaps similar to that of another nonsupersymmetric
case, namely Schwarzschild black holes (SBH) which are described by fundamental strings[\LEN-\DAM].
Unfortunately, this also means that calculations are not under control due to large corrections.

In ref. [\LEN,\SBH] the entropy of a $D$ dimensional SBH was obtained by viewing it as a highly excited string with a gravitationally
renormalized tension. The prescription used in ref. [\LEN,\SBH] is as follows. The near horizon geometry of a SBH is Rindler space with a dimensionless time and
energy. If one identifies the dimensionless Rindler energy with the (square root of the) string oscillator number then the entropy of the string
(the Rindler energy) gives the correct black hole entropy. The Hawking temperature is given by the Hagedorn temperature of the string which has a gravitationally
renormalized tension. For example, a $D=4$ SBH with mass $M$ is described by a string with a renormalized tension $T \sim (GM)^{-2}$ and a very high oscillator number
$n \sim E_R^2 \sim G^2 M^4$ (and mass $M$).
Another interpretation of these results is to think of the above string as living on the stretched horizon, a distance $\sqrt{\alpha^{\prime}}=\ell_s$ away from the
event horizon. On the stretched horizon the energy of the string with tension $T=1/2 \pi \alpha^{\prime}$ is $E_{sh} \sim G M^2/ \ell_s$. We see that for this string
$n \sim E_R^2$. However, an asymptotic
observer will see the gravitationally renormalized mass and tension given above.
In this case, the string describes the fundamental holographic degrees of freedom on the (stretched) horizon.
It is interesting to note that the above prescription works in other cases where the entropy of the system can be described by strings, e.g. near extreme
$D5$ branes [\JUA], some near extremal $D=4,5$ black holes[\EDI] and brane--antibrane systems[\DAN].

On the other hand, the near horizon geometry of de Sitter space is also given by Rindler space. This means that we can use the prescription of ref. [\LEN,\SBH]
to explain the entropy and temperature of de Sitter space in terms of a string with a renormalized tension.
It is clear that this prescription is very ad hoc and based on using the perturbative formulas for string entropy and mass in a very nonperturbative
setting, i.e. a highly excited and strongly interacting string. Nevertheless,
due to the success of the black hole entropy calculation and the
similarities between de Sitter and Schwarzschild space--times we will use the prescription of ref. [\LEN,\SBH] for the de Sitter case.

In this letter, we compute the Rindler energy near the de Sitter horizon and identify it with $\sqrt{n}$ for a string on the stretched horizon.
The string tension is renormalized
to  $T \sim \Lambda$ due to the large gravitational redshift near the horizon.
Then the string entropy
gives the (correct scaling of the) entropy of de Sitter space in any dimension. The temperature of de Sitter space is given by the Hagedorn temperature of the string
with the renormalized tension.
We take our results to be an indication that the entropy of the Sitter space is given by a very highly excited string with renormalized tension
which lives near the horizon.

This letter is organized as follows. In the next section we show that the prescription of ref. [\LEN,\SBH] gives the correct de Sitter entropy and
temperature in any dimension. In section 3, we discuss the validity of our results and try to interpret them.

\vfill

\medskip
\centerline{\bf 2. De Sitter Entropy and Strings}
\medskip

The $D$ dimensional Einstein--Hilbert action with a positive cosmological constant is
$$S={1 \over {16 \pi G_D}} \int d^Dx \sqrt{g}(R-2\Lambda) \eqno(1)$$
where  the cosmological constant is given by $\Lambda=1/L^2$ and $G_D$ is the $D$ dimensional Newton constant.
This action has the $D$ dimensional de Sitter space--time with the metric
$$ds^2=-dt^2+L^2cosh^2(t/L) d\Omega^2_{D-1} \eqno(2)$$
as a solution. Since there is no asymptotic spatial infinity in de Sitter space there is no notion of conserved energy[\WIT,\VJ]. However, an inertial observer
in de Sitter space sees a cosmological horizon. The static metric which describes the region inside the horizon is given by
$$ds^2=-\left(1-{r^2 \over L^2} \right)dt^2+\left(1-{r^2 \over L^2} \right)^{-1} dr^2+r^2 d\Omega_{D-2}^2 \eqno(3)$$
where $0 \leq r \leq L$. In these coordinates the horizon is at $r=L$. This static patch of de Sitter space has a globally defined time--like Killing vector
and therefore energy is well-defined in the patch[\VJ]. (From now on we will call the patch de Sitter space for brevity.)
The entropy of de Sitter space--time is given by the horizon area in Planck units
$$S_{dS}={A \over {4G_N}}={L^{D-2} A_{D-2} \over 4 G_D} \eqno(4)$$
where $A_{D-2}=2 \pi^{(D-1)/2}/\Gamma((D-1)/2)$ is the area of the $D-2$ dimensional unit sphere.
The temperature of de Sitter space--time (in any dimension) is
$$T_{dS}={1 \over {2 \pi L}} \eqno(5)$$

The near horizon geometry of de Sitter space is given by Rindler space. Near the horizon taking $r=L-y$ with $y<<L$ we find that the metric becomes
$$ds^2= -\left(2y \over L \right)dt^2+\left(2y \over L \right)^{-1} dy^2+L^2 d\Omega_{D-2}^2 \eqno(6)$$
The proper distance to the horizon is given by
$$R=\sqrt{L\over 8}\sqrt{y} \eqno(7)$$
Then the $t-R$ part of the metric becomes
$$ds^2=-\left(16 R^2 \over L^2 \right)dt^2+dR^2+\ldots \eqno(8)$$
which is the Rindler metric. The dimensionless Rindler time is
$$\tau_R={4 \over L} t \eqno(9)$$
The dimensionless Rindler energy $E_R$ is conjugate to the Rindler time, i.e. $[E_R,\tau_R]=1=(4/L)[E_R,t]$ which means
$$1={4 \over L} {\partial E_R \over \partial E} \eqno(10)$$
where the energy $E$ is conjugate to time $t$. For the SBH $E$ was the ADM mass of the black hole. The full de Sitter space is not asymptotically flat
and therefore there is no notion of ADM mass or conserved energy. However, there is a notion of energy for the static patch.
For $E$ we take the energy measured by a static observer at the center ($r=0$) of the static patch so that $E_{dS} \sim L^{(D-3)}/G_D$.
Note that this is also the mass of the maximal black hole that fits into de Sitter space with a given horizon size[\NAR].
This gives the Rindler energy
$$E_R \sim{L^{D-2} \over  G_D} \eqno(11)$$

Now, consider a string (with tension $T=1/2 \pi \alpha^{\prime}$) at the stretched horizon, $r=L-\ell_s$. The energy of this string is $E_{sh} \sim E_R/\ell_s$ and
therefore,
as in ref. [\LEN,\SBH], we identify the Rindler energy $E_R$ with the square root of oscillator number $\sqrt{n}$ of the string. Then we get
(This formula is only valid for perturbative strings however we will use it in this highly nonperturbative situation.)
$$S_{str}=2 \pi \sqrt {c/6} E_R \sim S_{dS}\eqno(12)$$

We find that the entropy of de Sitter space in any dimension is given by the entropy of a string near the horizon with oscillator number $n \sim E_R^2$.
This is a very long string of length $\sim E_R \ell_s$. The entropy of the string is
proportional to its length. This can be explained if we assume that each string bit of length $\ell_s$ carries one bit of information. The string bits are
the fundamental holographic degrees of freedom on the horizon with a density of one per Planck area.

On the stretched horizon, the energy of the string is $E_{sh} \sim L^{(D-2)}/ \ell_s G_D$. However,
the energy seen by an observer at the center of de Sitter space (or that of the largest possible black hole that fits inside the horizon)
is given by
$$E_{dS} \sim {L^{D-3} \over G_D} \sim {\sqrt{n \over T}} \sim {E_R \over \sqrt{T}}  \eqno(13)$$
where we identified the total energy seen at $r=0$ with the energy of the string (where once again we use the perturbative formula).
We find that the tension of
the string is renormalized due to the large redshift near the horizon from $T=1/ 2 \pi \alpha^{\prime}$ to $T  \sim 1/L^2 = \Lambda$. Thus, the Hagedorn temperature of
the string near the horizon (in any dimension) becomes
$$T_{Hag} \sim {1 \over {2 \pi L}} \eqno(14)$$
which is the temperature of de Sitter space in any dimension. Following ref. [\HRS] one can also show that the above string will emit black body radiation
(small closed strings) at the de Sitter temperature.
The de Sitter temperature can also be obtained by noting that the Rindler temperature is $T_R =1/2 \pi$. Using eq. (9) we find that a static observer at $r=0$
will see a temperature $T_{dS} \sim 1/ 2 \pi L$.

\vfill

\centerline{\bf 3. Conclusions and Discussion}
\medskip

We showed that the entropy and temperature of de Sitter space in any dimension can be understood as that of a very highly excited string
with a renormalized tension located near the de Sitter horizon. Actually, we were only able to obtain the correct scaling of the entropy
up to dimension dependent factors. It is not clear how to improve the prescription used above to get the correct numerical factors. For example,
if one could vary the cosmological constant without changing the radius of the horizon this would give the numerically correct entropy in all dimensions.
However, this does not seem to be physical. In any case, due to the lack of supersymmetry there must be large corrections to all calculations
(unless the above prescription takes into account all such corrections) which
make results about numerical factors irrelevant.

As mentioned above, in order to obtain this result we used the perturbative string formulas
for this very nonperturbative and strongly interacting string without any justification (except the fact that they give the correct results for
the SBH and de Sitter space). In addition, we used for $E$ the total energy inside the horizon arising from the positive cosmological constant in de Sitter
space. This energy is
also the mass of the largest black hole that fits into the de Sitter horizon. The interpretation of our results depends on which (if either) one of
these origins of $E$ is the correct one. For example, if the maximal black hole mass is the relevant quantity, one may imagine that these largest
possible (virtual) black holes play an important role in describing physics in de Sitter space.

It is amusing that the fundamental holographic degrees of freedom of de Sitter may be given by (maybe noncritical) strings (with renormalized tension) even
though we cannot formulate string theory as we understand it in bulk of de Sitter space.



\vfill

\refout

\end
\bye